\documentclass{PoS}

\def\lsi{\raise0.3ex\hbox{$<$\kern-0.75em\raise-1.1ex\hbox{$\sim$}}}
\def\gsi{\raise0.3ex\hbox{$>$\kern-0.75em\raise-1.1ex\hbox{$\sim$}}}
\def\backder{\raise1.4ex\hbox{$\leftarrow$\kern-0.75em\raise-1.4ex\hbox{$\partial$}}}
\def\forder{\raise1.4ex\hbox{$\rightarrow$\kern-0.8em\raise-1.4ex\hbox{$\partial$}}}

\newcommand{\gsim}{\mathop{\gsi}}
\newcommand{\C}{{\kern+.25em\sf{C}\kern-.45em\sf{{\small{I}}} \kern+.45em\kern-.25em}}
\newcommand{\R}{{\kern+.25em\sf{R}\kern-.78em\sf{I} \kern+.78em\kern-.25em}}
\newcommand{\N}{{\kern+.25em\sf{N}\kern-.78em\sf{I} \kern+.78em\kern-.25em}}
\newcommand{\beq}{\begin{equation}}
\newcommand{\eeq}{\end{equation}}
\newcommand{\bea}{\begin{eqnarray}}
\newcommand{\eea}{\end{eqnarray}}
\newcommand{\nn}{\nonumber}
\newcommand{\la}{\langle}
\newcommand{\ra}{\rangle}

\newcommand{\uno}{1 \!\!\! 1}

\title{Simulations of a supersymmetry inspired \\ model on a fuzzy sphere}

\ShortTitle{SUSY an a fuzzy sphere}

\author{Wolfgang Bietenholz
\thanks{Preprint \ \ DESY-07-123, published in PoS(LAT2007)283.} \\ 
\ \\
John von Neumann Institut f\"{u}r Computing NIC \\
Deutsches Elektron-Synchrotron DESY \\
Platanenallee 6 \\
D-15738 Zeuthen, Germany \\
\ \\        
E-mail: \email{bietenho@ifh.de \\ }}

\abstract{We present a numerical study of a two dimensional
model of the Wess-Zumino type. We formulate this model on a 
sphere, where the fields are expanded in spherical harmonics.
The sphere becomes fuzzy by a truncation in the angular momenta.
This leads to a finite set of degrees of freedom without explicitly
breaking the space symmetries. The corresponding field theory
is expressed in terms of a matrix model, which can be simulated. 
We present first numerical results for the phase structure
of a variant of this model on a fuzzy sphere. The prospect to restore
exact supersymmetry in certain limits is under investigation.}

\FullConference{The XXV International Symposium on Lattice Field
Theory\\
                 July 30 - August 4 2007 \\
                 Regensburg, Germany}

\begin{document}

\section{The Di Vecchia-Ferrara model}

Back in 1977 Di Vecchia and Ferrara introduced an elegant supersymmetric
(SUSY) model of the Wess-Zumino type in two dimensions \cite{divefe}. 
On a Euclidean plane its action reads
\beq
S [ \phi, \psi ] = \frac{1}{2} \int d^{2}x \, \Big[ \partial_{\mu}\phi
\partial_{\mu}\phi + \bar \psi \gamma_{\nu} \partial_{\nu}
\psi + [ V'(\phi)]^{2} + \bar \psi \psi \, V''(\phi ) \Big] \ ,
\eeq
where $\phi$ is a scalar field, and $\psi$ is a 2-component
Majorana spinor field (with $\bar \psi = \psi^{T} C$, where
$C=i \sigma_{2}$ is the charge conjugation operator).

We fix the boson-fermion interaction, as well as the bosonic potential,
by the choice
\bea
V( \phi ) &=& \sqrt{ \frac{\lambda}{108}} \phi^{3} + \frac{m}{2} \phi^{2} \\
\Rightarrow \quad \frac{1}{2} [ V'(\phi )]^{2} &=&
\frac{\lambda}{4!} \phi^{4} + m \sqrt{\frac{\lambda}{12}} \phi^{3}
+ \frac{m^{2}}{2} \phi^{2} \ , \quad
\frac{1}{2} V''(\phi ) = \sqrt{\frac{\lambda}{12}} \phi + m \nn \ .
\eea
This action is invariant (up to a total divergence) under the
SUSY transformation
\beq
\left\{ \begin{array}{ll}
\delta \phi = \bar \epsilon \psi & \\
\delta \psi = [ \gamma_{\nu} \partial_{\nu} \phi - V'(\phi ) ] \epsilon
& , \quad \delta \bar \psi = \bar \epsilon
[ - \gamma_{\nu} \partial_{\nu} \phi - V'(\phi ) ] \ ,
\end{array} \right.
\eeq
where $\epsilon$ is a constant spinor field.

\section{The fuzzy sphere}

\subsection{Geometry}

The coordinate operators $\hat x_{i}$ on a fuzzy sphere of radius $R$
have to solve the equation \cite{Madore}
\beq
\hat x_{1}^{2} + \hat x_{2}^{2} + \hat x_{3}^{2} = R^{2} \uno \ .
\eeq
This can be achieved by identifying the position operators 
$\hat x_{i}$ with the (re-scaled) angular momentum operators $\hat L_{i}\,$, 
\beq
\hat x_{i} = \frac{R}{\sqrt{\ell ( \ell +1)}} \, \hat L_{i} \ ,
\qquad i = 1, \dots , 3 \ , \quad \ell \in \N \ ,
\eeq
where $\ell$ is the spin of the irreducible representation.
For finite $\ell$ we obtain $N \times N$ matrices, with $N = \ell +1 \,$.
Then the coordinate operators on the fuzzy sphere are non-commutative,
\beq
[ \hat x_{i} , \hat x_{j} ] = {\rm i} \frac{R}{\sqrt{\ell ( \ell +1)}}
\epsilon_{ijk} \hat x_{k} \ .
\eeq
In the limit $\ell \to \infty$ commutativity is restored, and the
sphere is not fuzzy anymore.

\subsection{Fields}

Some field, for instance our scalar field $\phi (x)$,
on the sphere $S^{2}$ can be expanded in term of spherical harmonics
$Y_{\ell m}(x) \, $,
\beq  \label{expand1}
\phi (x) = \sum_{\ell =0}^{\infty} \, \sum_{m=-\ell}^{\ell}
c_{\ell m} Y_{\ell m}(x) \ .
\eeq

In accordance with the above treatment of the coordinates,
we are going to represent also the fields as matrices, so
that we end up with a matrix model.
Without limiting the matrix size $N$, a Hermitian
matrix $\hat \phi$ can be expanded in the {\em polarisation tensors} 
$\hat Y_{\ell m}$ \cite{polten} , in analogy to eq.\ (\ref{expand1}),
\beq  \label{matexpand}
\hat \phi = \sum_{\ell =0}^{\infty} \, \sum_{m=-\ell}^{\ell}
c_{\ell m} \hat Y_{\ell m} \ .
\eeq
The polarisation tensors play a r\^{o}le analogous to the spherical 
harmonics. In particular they obey $\hat L^{2}
\hat Y_{\ell m} = \ell (\ell +1) \hat Y_{\ell m} \, $, where
$\hat L^{2} $
is the angular momentum squared (cf.\ Section 3).
Moreover we have the adequate normalisation and parity behaviour,
$\frac{4 \pi}{N} {\rm Tr} ( \hat Y_{\ell' m'}^{\dagger}
\hat Y_{\ell m} ) = \delta_{\ell ' \ell} \delta_{m' m} \ , \ \
\hat Y_{\ell m}^{\dagger} = (-1)^{m} \hat Y_{\ell -m} \ $.

We now introduce a cutoff $\ell_{\rm max} = N-1$, and the remaining 
coefficients for the field $\phi$
can be embedded into Hermitian $N \times N$ matrices.
Thus we arrive at a finite set of degrees of freedom, 
without any explicit breaking of the space symmetries.
Therefore this regularisation is attractive for
SUSY models, where the lattice formulation is notoriously 
troublesome \cite{LatSUSY}. Moreover the fuzzy sphere regularisation 
is not plagued by the fermion doubling problem \cite{BalIm}.

Theoretical aspects of SUSY on a fuzzy sphere
are discussed in Refs.\ \cite{FuzzySUSY} and reviewed in 
Ref.\ \cite{Bal_lectures}.
The feasibility of numerical simulations in
this regularisation has been tested for the $\lambda \phi^{4}$
model in $d=2$ \cite{Xavier} and $d=3$ \cite{Julieta}.
For an approach to simulate SUSY gauge theory on the fuzzy sphere,
see Ref.\ \cite{FuzzySUSYnum}. 

\section{The Di Vecchia-Ferrara model on a fuzzy sphere}

To be explicit, we transfer the Di Vecchia-Ferrara model 
from the Euclidean plane to a fuzzy sphere by means of the
following substitutions:
\bea
\phi (x) \ \to \ \hat \phi & \  , \quad & 
\frac{1}{4 \pi R^{2}} \int_{|x|=R} d \Omega \,
\phi (x) \ \to \ \frac{1}{N} {\rm Tr} ( \hat \phi ) \ , \nn \\
\partial_{i} \phi (x) \ \to \ \hat \partial_{i} \hat \phi =
{\rm i} [ \hat L_{i} , \hat \phi ] 
& \ , \quad &
- \partial^{2} \phi (x) \ \to \ \sum_{i=1}^{3} 
[ \hat L_{i} , [ \hat L_{i}, \hat \phi ]] \ ,  \label{translate}
\eea 
where $\hat \phi$ is a Hermitian $N\times N$ matrix.
We are ultimately interested in the limits $N , \ R \to \infty$.

In practice the left-handed and right-handed applications of the
operators 
can be implemented
best by storing the matrix configurations $\hat \phi$
as vectors. In this setting the Dirac operator (\ref{Diracop})
takes the form of a $2 N^{2} \times 2 N^{2}$ matrix.
We symmetrise the fermionic potential as
\beq
V''_{\rm sym} (\hat \phi ) = \frac{1}{2} \Big( V''(\hat \phi ) \otimes
\uno_{N} + \uno_{N} \otimes V''(\hat \phi ) \Big) \ ,
\eeq
which leads to the Dirac operator
\beq  \label{Diracop}
D = \left( \begin{array}{cc}
\hat \partial_{3} + \frac{\rm i}{R} \uno_{N}
+ V''_{\rm sym} & \hat \partial_{-} \\
\hat \partial_{+} & - \hat \partial_{3} + \frac{\rm i}{R}
\uno_{N} + V''_{\rm sym}
\end{array} \right)
\eeq
with $\hat \partial_{\pm} = 
\hat \partial_{1} \pm {\rm i} \hat \partial_{2} \, $, and with the
representation (\ref{translate}) of the differential operators.
As usual in Euclidean space we deal with an anti-Hermitian
kinetic part. This also includes the
term ${\rm i}/R \, $, which emerges from the curvature
effect on the spin connection.
 
Integrating out the fermionic variables yields the Pfaffian
${\rm Pf} \, [(CD)_{\rm asym}]$, where the subscript means the
anti-symmetric part. As a first approach to explore this type of
model, we replace the Pfaffian by $\sqrt{{\rm det} \, D} \, $, 
\footnote{The impact of this substitution remains to be investigated
in detail. Further comments are added in Section 5.}
so we arrive at the matrix model action
\beq  \label{fullact}
S [\phi ] = \frac{4\pi}{N} {\rm Tr} \Big(
\frac{1}{2} \sum_{i=1}^{3} [ \hat L_{i} , [ \hat L_{i}, \hat \phi ]]
+ \frac{m^{2}}{2} \hat \phi^{2} + m \sqrt{\frac{\lambda}{12}} \hat \phi^{3}
+ \frac{\lambda}{4!} \hat \phi^{4} \Big) 
- \ln \sqrt{{\rm det} \, D} \ .
\eeq

\section{Order parameters}

As in Refs.\ \cite{Xavier,Julieta} we are going to
explore the phase diagram by considering order parameters, which
are constructed from the coefficients $c_{\ell m}$ in the
expansion (\ref{matexpand}). They can be extracted from the
relation
\beq
c_{\ell m} = \frac{4\pi}{N} {\rm Tr} \,
[ ( \hat Y_{\ell m})^{\dagger} \hat \phi ] \ .
\eeq
In particular we focus on the quantities
\beq
\phi_{\ell}^{2} := \sum_{m=-\ell}^{\ell} | c_{\ell m}|^{2} \qquad 
{\rm and} \qquad
| \phi |^{2} := \sum_{\ell} \phi_{\ell}^{2} =
\frac{4\pi}{N} {\rm Tr} \, [ \hat \phi^{2} ] \ .
\eeq
Based on the magnitudes of the expectation values
$\la \phi_{0}^{2} \ra$ and $ \la | \phi |^{2} \ra$ we distinguish
three phases as specified in Table \ref{phasetab},
in close analogy to the $\lambda \phi^{4}$ model on
a non-commutative flat space \cite{NCphi4}:
\begin{table}
\begin{center}
\begin{tabular}{|c|c|c|}
\hline
phase &  $\la \phi_{0}^{2} \ra$ & $ \la | \phi |^{2} \ra$ \\
\hline
\hline
disordered         &  $\approx 0 $ & $\approx 0 $ \\
\hline
uniform ordered    &  $\gg 0 $ & $\gg 0 $ \\
\hline
non-uniform ordered & $\approx 0 $ & $\gg 0 $ \\
\hline
\end{tabular}
\vspace*{2mm}
\caption{\emph{The phases that we observed,
along with the respective magnitudes of the order parameters.}}
\label{phasetab}
\end{center}
\vspace*{-2mm}
\end{table}

\begin{itemize}

\item In the {\em disordered phase} $\la \phi_{\ell}^{2} \ra \approx 0$
holds for all $\ell$. The angular mode decomposition does not detect
any contribution that could indicate spontaneous symmetry breaking.

\item The {\em uniform ordered phase} is characterised by
$ \la | \phi |^{2} \ra \approx \la \phi_{0}^{2} \ra \gg 0\,$,
{\it i.e.}\ the zero mode contributes significantly, whereas
higher modes are suppressed. This phase corresponds
to the spontaneous magnetisation in an Ising-type spin model.

\item In the {\em non-uniform ordered phase} a non-zero mode
condenses. This leads to the
relations $ \la | \phi |^{2} \ra \approx 
\la | \phi | ^{2} - \phi_{0}^{2} \ra \gg 0\,$. 
In this case the rotation symmetry is spontaneously broken.
That phase is specific to the {\em fuzzy} sphere; it does not
occur in commutative spaces.

\end{itemize}

We add that the fuzzy sphere formulation is non-local,
as the non-commutativity of the coordinates suggests.
Therefore the Mermin-Wagner theorem does not prohibit
the non-uniform ordered phase.

\section{Numerical results for the phase diagram}

In general the determinant ${\rm det} \, D$ (where $D$ is given
in eq.\ (\ref{Diracop})) is complex. In the final limit
$N , \, R \to \infty$ it is real positive, however, hence the
complex phase represents an artifact of the fuzzy
regularisation (apart from the substitution in eq.\ (\ref{fullact})).
We therefore modify the regularisation at
this point by using $|{\rm det} \, D|$ already at finite
$N$ and $R$.
With this modification the action (\ref{fullact}) defines
a Boltzmann weight which enables Monte Carlo simulations.
We performed such simulations with the Metropolis algorithm:
in each step, a conjugate pair of matrix elements is updated, 
and $|{\rm det} \, D|$ is computed explicitly.
Throughout our simulations we fixed the radius of the
sphere to $R=1$.

\begin{figure}[h!]
\vspace*{-3mm}
\begin{center}
\includegraphics[angle=270,width=.49\linewidth]{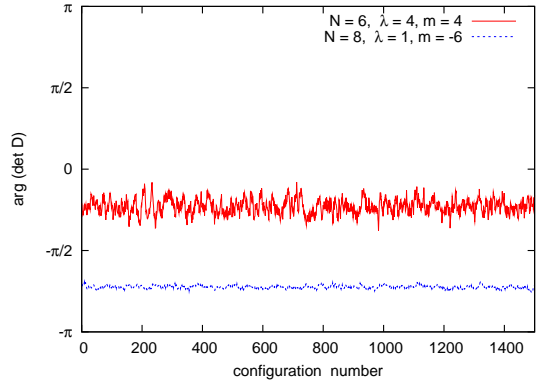}
\vspace*{-2mm}
\end{center}
\caption{{\it Two examples for histories of the phase of the
fermion determinant. The parameters in these simulations
are $N=6, \ \lambda = m =4$ and $N=8, \ \lambda = 1, \, m = -6 \,$.}}
\label{phasehisto}
\vspace*{-1mm}
\end{figure}

So far we simulated at $N=4$, 6 and 8
(which are numerically handled by matrices of size 32, 72 and 128)
and we explored the
phase diagram in the $(m, \lambda)$ plane. We also
measured the phase of the determinant. It turns out to be quite
stable, which is favourable for the modified regularisation
(in the extreme case of an invariant phase the modification 
is redundant). Examples for this property are illustrated
in Fig.\ \ref{phasehisto}.
 
\begin{figure}[h!]
\begin{center}
\includegraphics[angle=270,width=.32\linewidth]{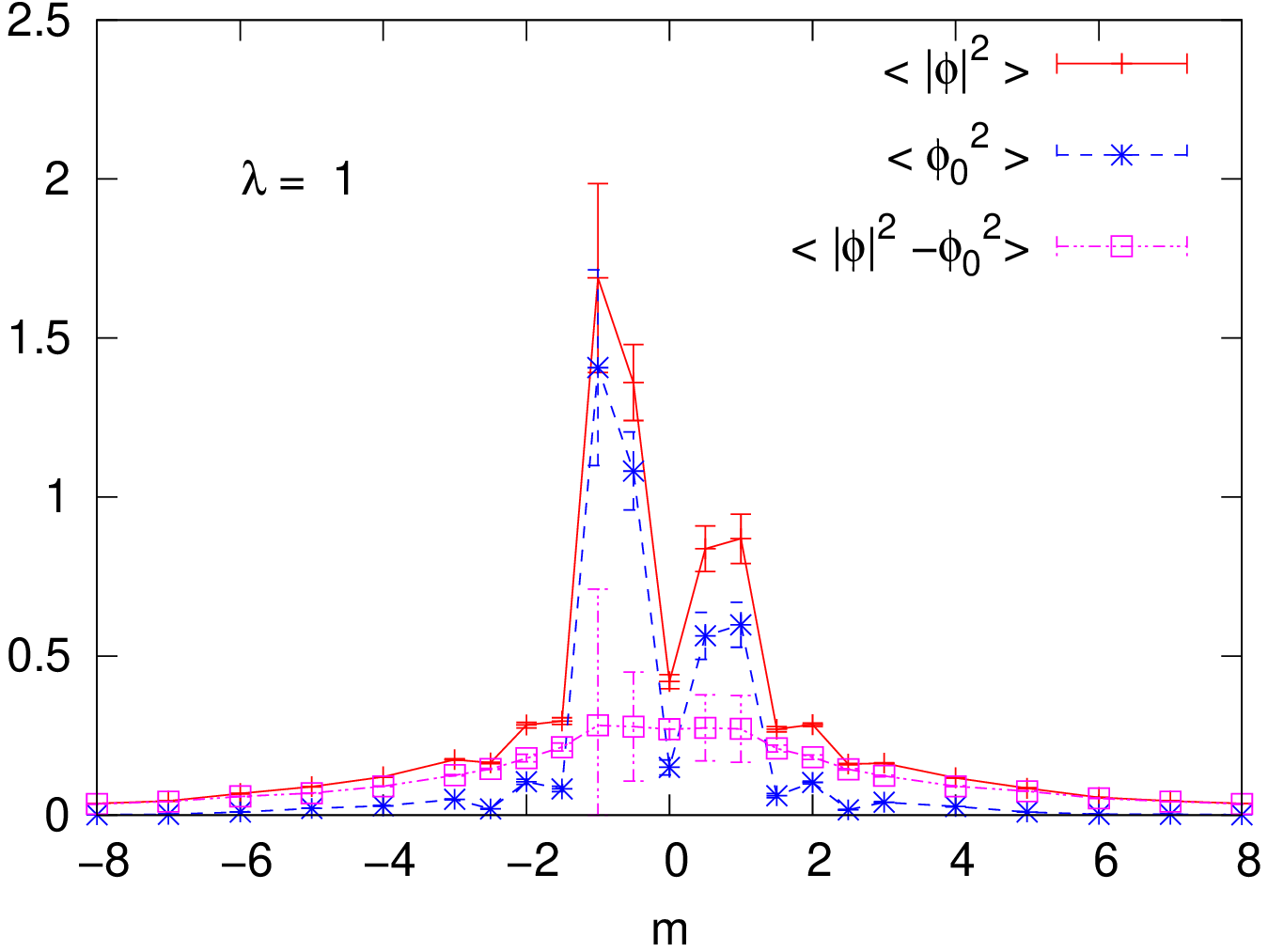}
\includegraphics[angle=270,width=.32\linewidth]{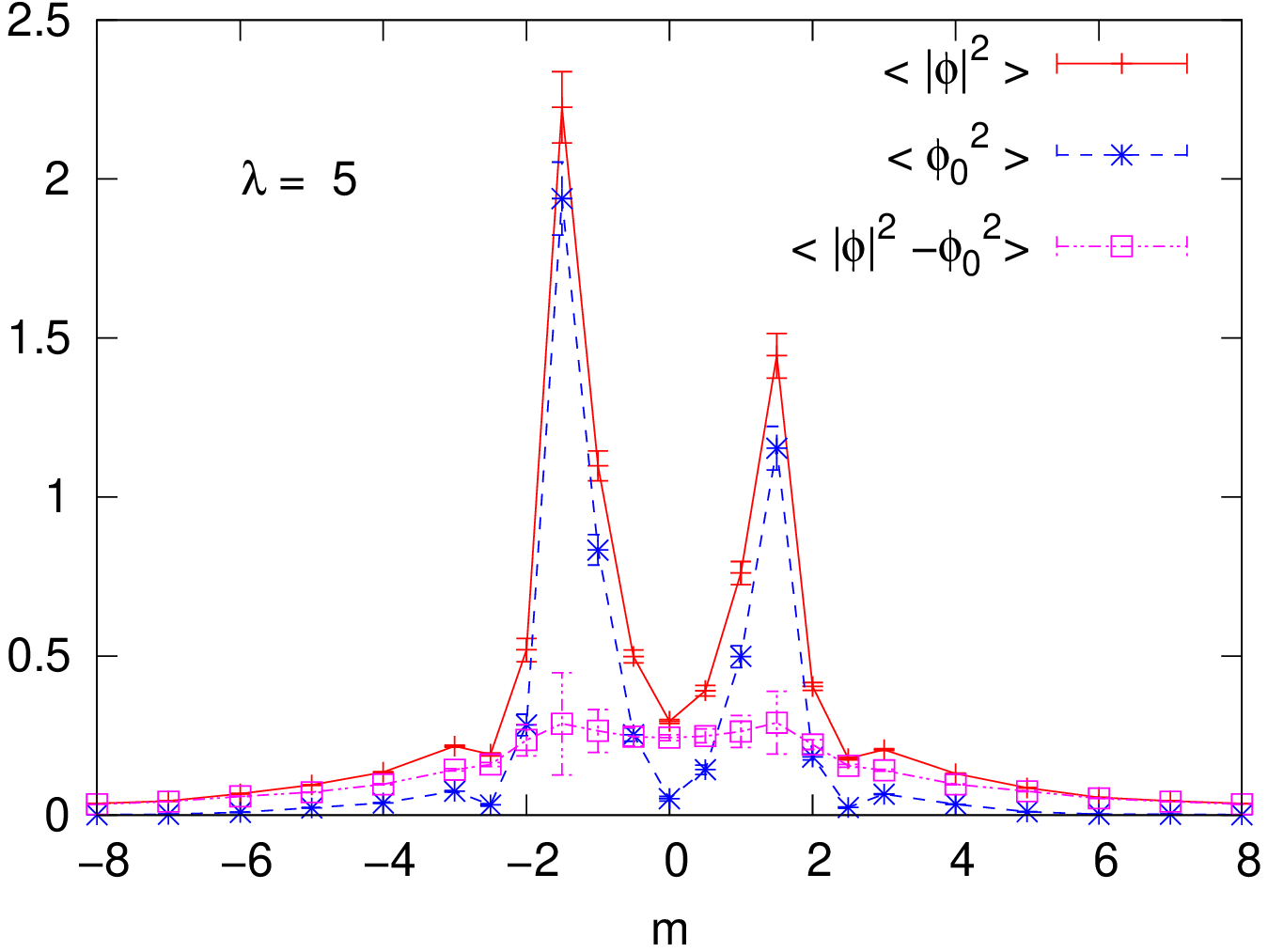}
\includegraphics[angle=270,width=.32\linewidth]{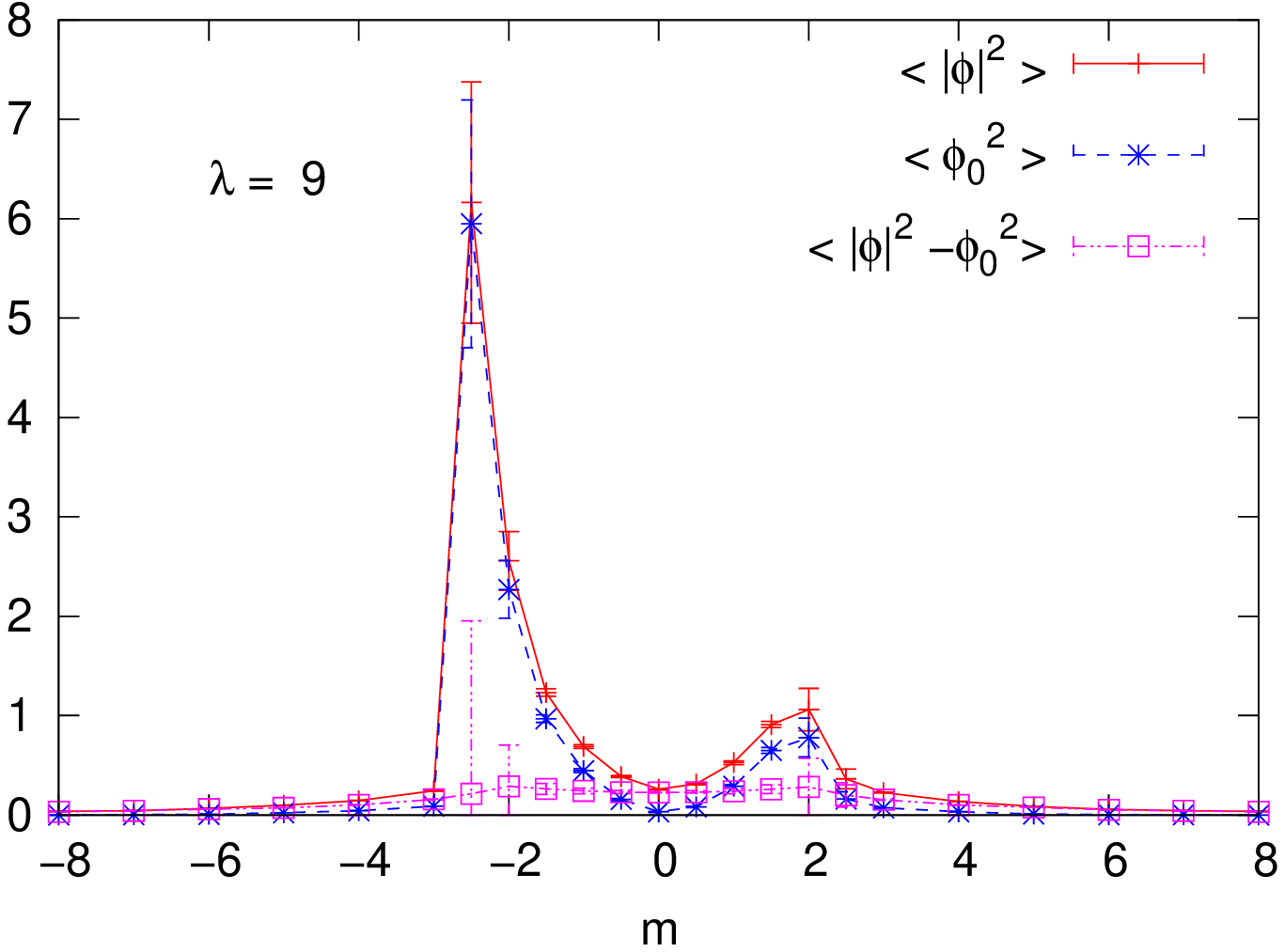}
\end{center}
\caption{{\it The order parameters for $N=6$ and $ \lambda = 1,\, 5,\, 9$
in the range $m = -8 \dots 8$. We see a double peak structure
in all cases, and the peak at $m \approx -2$ is enhanced for
increasing $\lambda$. A strong coupling $\lambda$
also implies a gradually enlarged phase of non-uniform order.}}
\label{N6orpa}
\end{figure}

\begin{figure}[h!]
\begin{center}
\includegraphics[angle=270,width=.32\linewidth]{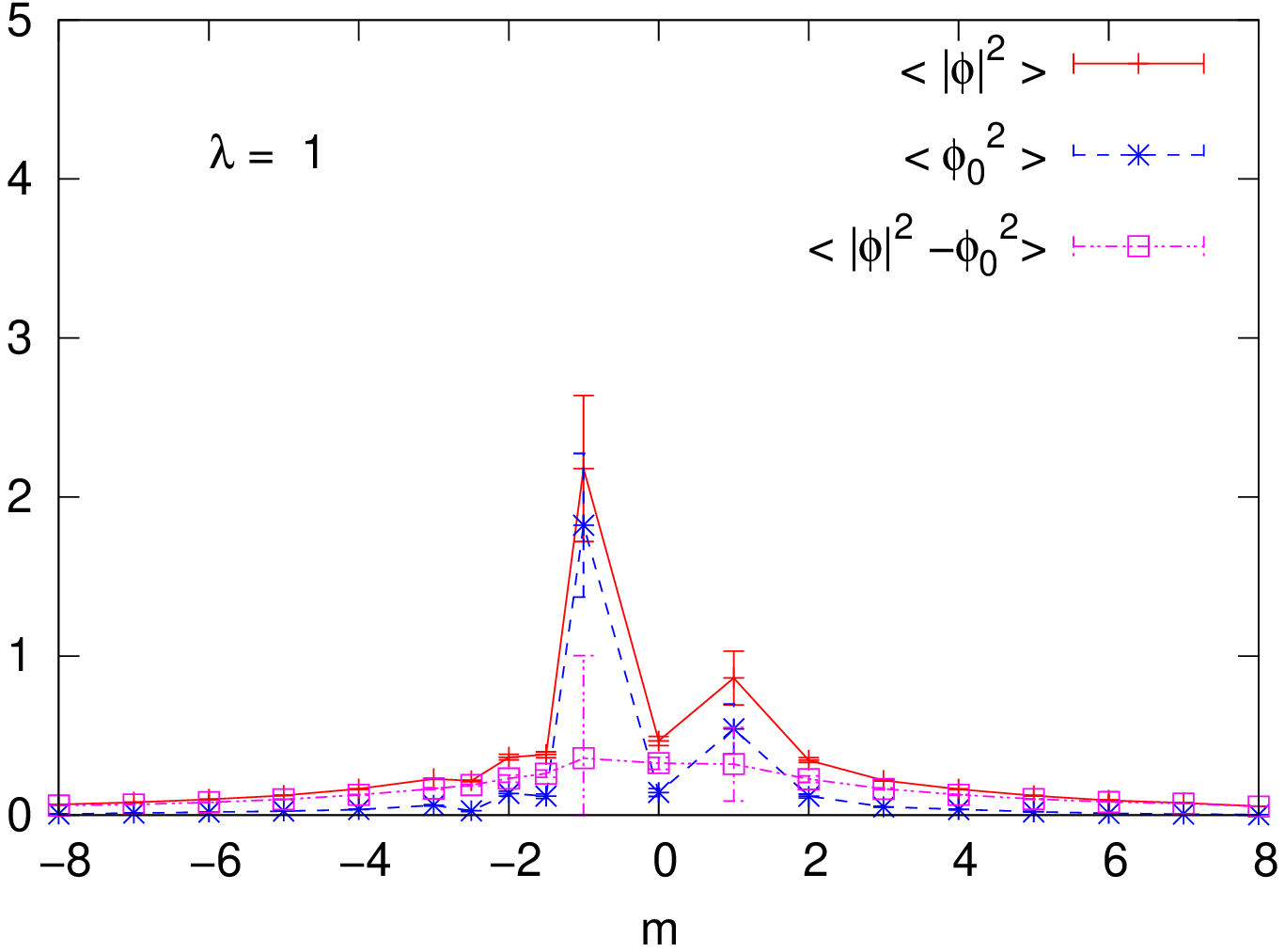}
\includegraphics[angle=270,width=.32\linewidth]{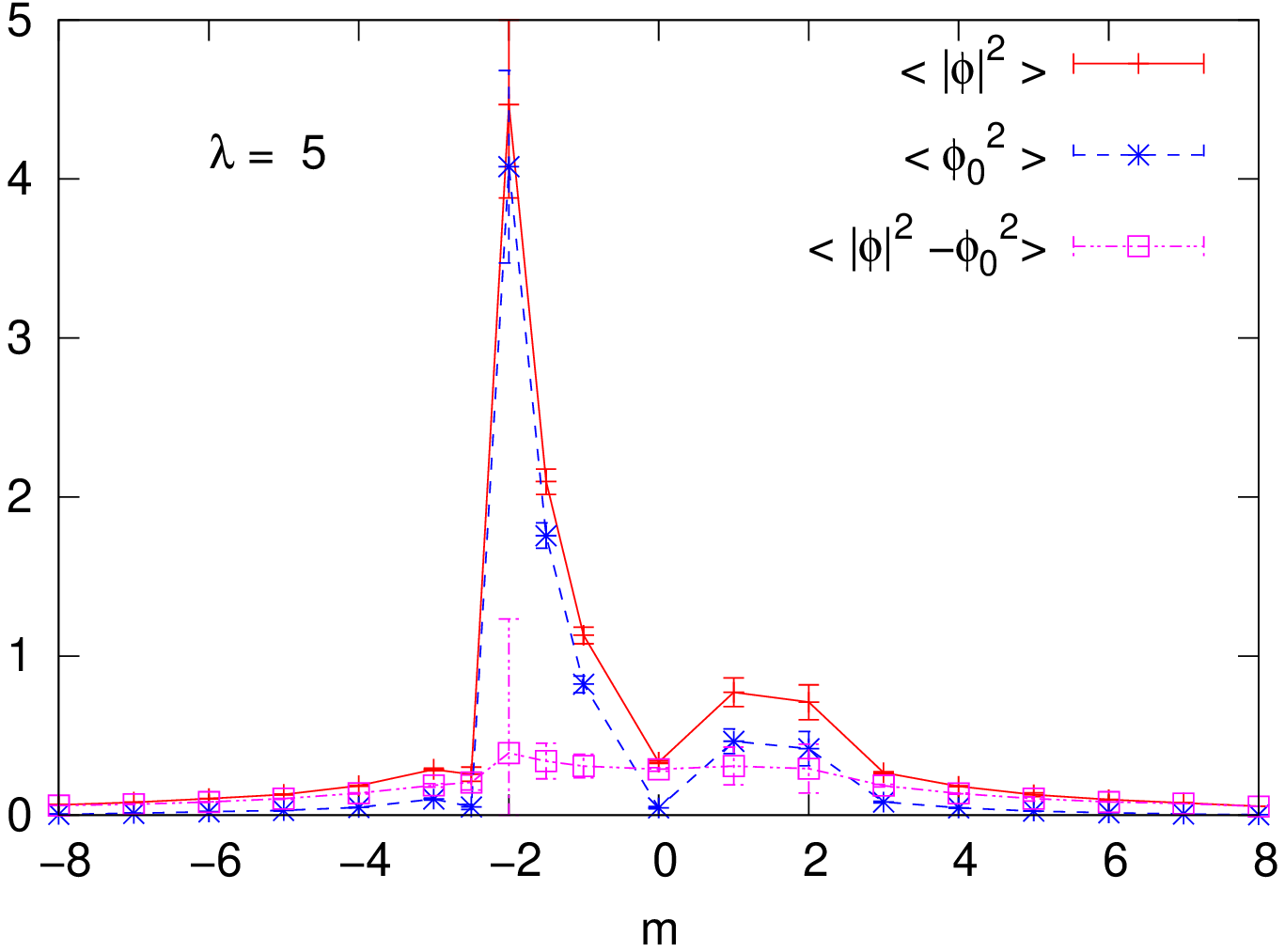}
\includegraphics[angle=270,width=.32\linewidth]{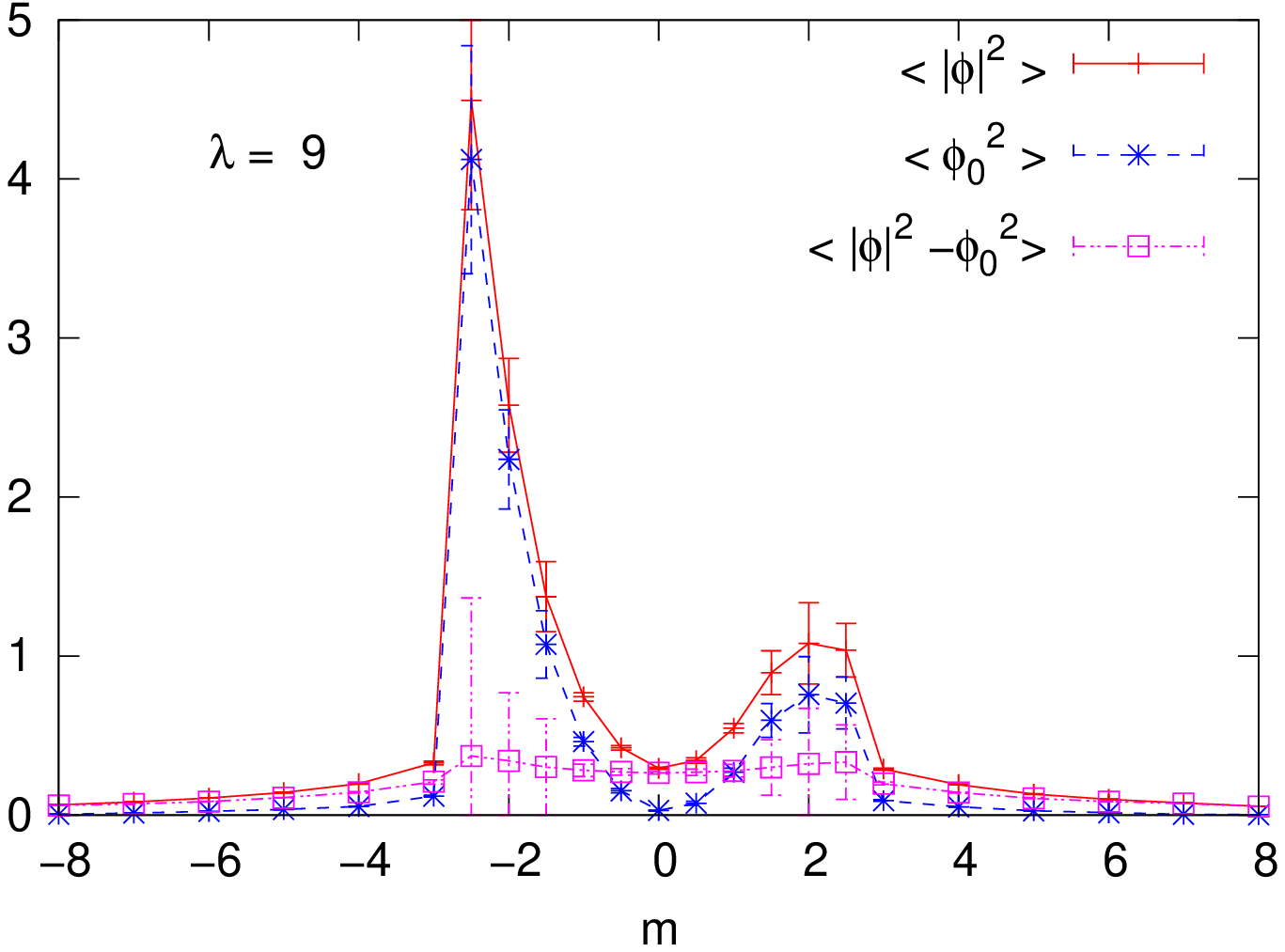}
\end{center}
\caption{{\it The order parameters for $N=8$ and $ \lambda = 1,\, 5,\, 9$
in the range $m = -8 \dots 8$. The phase transitions show a
behaviour similar to the results for $N=6$ (Fig.\ 2), although
the values of the order parameters differ.}}
\label{N8orpa}
\end{figure}

Next we show in Figs.\ \ref{N6orpa} and \ref{N8orpa}
the expectation values of the order parameters
$\langle | \phi |^{2} \rangle$, $\langle \phi_{0}^{2} \rangle$ and
$\langle | \phi |^{2} - \phi_{0}^{2} \rangle$ (as discussed in
Section 4) 
for $N=6$ and $N=8$, at $\lambda = 1, \, 5$ and $9\, $.
We actually investigated a larger range of the mass parameter $m \,$, but
we show here the interval of interest in view of the phase diagram.
In all cases, large values of $| m |$ lead to the disordered
phase. When $|m|$ decreases below a critical value $\approx 2$ 
(which is similar but not identical for both signs of $m$) 
we enter the phase of uniform order.
In the vicinity of $m = 0$ we also observe the phase of 
non-uniform order to set in; the corresponding conditions are
given in Table \ref{phasetab}.


\begin{figure}[h!]
\begin{center}
\includegraphics[angle=270,width=.5\linewidth]{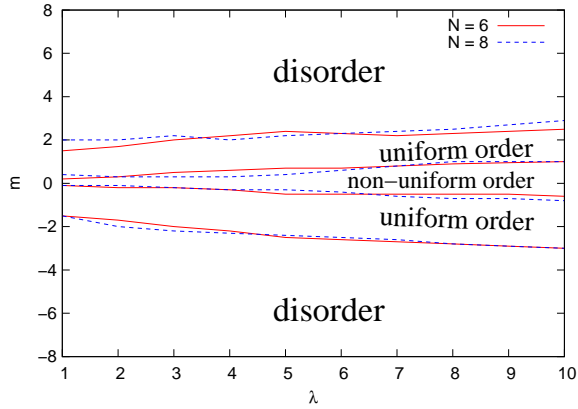}
\end{center}
\caption{{\it The phase diagram at $N=6$ and $8$, which is identified 
from measurements as shown in Figs.\ 2 and 3 for 
$\lambda =1,\, 2,\ \dots , 10$.}}
\label{phasediaN6}
\vspace*{-3mm}
\end{figure}

For $N=6$ and 8 we probed $\lambda = 1, \dots , 10\,$, 
which gives rise to the phase diagram in 
Fig.\ \ref{phasediaN6}.

\section{Conclusions}

We explored a new way to simulate a two dimensional model of the
Wess-Zumino type. The model is wrapped on a sphere and
the fields are expanded in spherical harmonics. A truncation in the
angular momentum renders the sphere fuzzy, and the corresponding
field coefficients build a finite set of degrees of freedom to be
used in numerical simulations. In this first approach we simplified
the Pfaffian to $\sqrt{{\rm det} D}$. Thus we studied a SUSY inspired
system of interacting scalars and Majorana fermions on a fuzzy sphere.

In the final limit of infinite angular momentum cutoff $N$ and 
radius $R$, the determinant ${\rm det} D$ is real positive.
This does not hold at finite
$N$ and $R$, so we modify the regularisation by employing
the modulus of the fermion determinant already on the regularised
level. We expect this formulation to lead to the same 
limit. This expectation is supported by the observation that
the fluctuations of the phase $\, {\rm arg} ({\rm det} D)\,$ are small.

With this method we simulated the system at $N=4$, $6$ and $8$ 
and $R=1$. The basic properties are similar in all cases; in 
particular large $|m| \gsim 2$ always leads to a disordered phase.
So it is conceivable that we are already peeping at aspects of the
large $N$ limit. However, the final stabilisation
of the phase diagram at large $N$ may involve a re-scaling of the axes. 

The ordered non-uniform phase emerges as a consequence of the 
non-commutativity of the coordinates, which we use on the regularised level.
We observed that phase around $m \approx 0$, and it ought to 
evaporate as we proceed to larger $N$. 
That feature is left for further investigation.

Although this project is still on-going, the preliminary
results are encouraging regarding the hope to find a way
to formulate and explore SUSY inspired models beyond perturbation 
theory. \\

\vspace*{-3mm}

\noindent
{\small
{\bf Acknowledgements :} This work is based on collaboration
with D.\ O'Connor, M.\ Panero and J.\ Volkholz. 
J.V.\ presented this talk, but meanwhile he left physics and
he preferred to withdraw his name.
I am also indebted to A.\ Balachandran, J.\ Medina, A.\ Wipf and B.\ Ydri 
for helpful discussions.
Most computations were performed on 
clusters of the ``Norddeutscher Verbund f\"ur Hoch- und 
H\"ochstleistungsrechnen'' (HLRN).
}

\end{document}